\documentclass[11pt]{article}
\usepackage{graphicx}
\usepackage{amsmath}
\usepackage{amssymb}
\usepackage{caption2}
\begin{document}
\bibliographystyle{prsty}
\begin{center}
{\large {\bf \sc{ Analysis of strong decays of the charmed mesons $D(2550)$, $D(2600)$, $D(2750)$ and $D(2760)$}}} \\[2mm]
Zhi-Gang Wang  \footnote{E-mail:wangzgyiti@yahoo.com.cn. } \\
  Department of Physics, North China Electric Power University, Baoding 071003, P. R.
  China
\end{center}

\begin{abstract}
In this article, we study the strong decays of  the newly observed
charmed mesons $D(2550)$, $D(2600)$, $D(2750)$ and $D(2760)$ with
the heavy quark effective theory in the leading order approximation,
and tentatively identify the $(D(2550),D(2600))$ as the $2S$ doublet
$(0^-,1^-)$ and the $(D(2750),D(2760))$  as the $1D$  doublet
$(2^-,3^-)$, respectively. The identification of the $D(2750)$ and
$D(2760)$ as the same particle with $J^P=3^-$ is disfavored.
\end{abstract}


\section{Introduction}

Recently the Babar collaboration observed four excited charmed
mesons $D(2550)$, $D(2600)$, $D(2750)$ and $D(2760)$ in the decay
channels $D^0(2550)\to D^{*+}\pi^-$, $D^{0}(2600)\to
D^{*+}\pi^-,\,D^{+}\pi^-$,  $D^0(2750)\to D^{*+}\pi^-$,
$D^{0}(2760)\to D^{+}\pi^-$, $D^{+}(2600)\to D^{0}\pi^+$ and
$D^{+}(2760)\to D^{0}\pi^+$ respectively in the inclusive $e^+e^-
\rightarrow c\bar{c}$  interactions  at the SLAC PEP-II
asymmetric-energy collider \cite{Babar2010}, see Table 1.  The Babar
collaboration also analyzed  the helicity distributions to determine
the spin-parity, and suggested that the $(D(2550),D(2600))$ (denoted
as $(D^{\prime},D^{*\prime})$ respectively in Table 2) may be
 the $2S$ radial excitation of the $(D,D^*)$, and the
$D(2750)$ and $D(2760)$ may be the $D$-wave states. Furthermore, the
Babar collaboration measured the following  ratios of the branching
fractions:
\begin{eqnarray}
\frac{{\rm{Br}}\left(D^*_2(2460)^0\to
D^+\pi^-\right)}{{\rm{Br}}\left(D^*_2(2460)^0\to
D^{*+}\pi^-\right)}&=&1.47\pm0.03\pm0.16 \, , \nonumber \\
\frac{{\rm{Br}}\left(D(2600)^0\to
D^+\pi^-\right)}{{\rm{Br}}\left(D(2600)^0\to
D^{*+}\pi^-\right)}&=&0.32\pm0.02\pm0.09\, , \nonumber \\
\frac{{\rm{Br}}\left(D(2760)^0\to
D^+\pi^-\right)}{{\rm{Br}}\left(D(2750)^0\to
D^{*+}\pi^-\right)}&=&0.42\pm0.05\pm0.11\, .
\end{eqnarray}

In the heavy quark limit $m_Q \to \infty$, the heavy-light  mesons
$Q{\bar q}$  can be  classified in doublets according to the total
angular momentum of the light degrees of freedom ${\vec s}_\ell$,
${\vec s}_\ell= {\vec s}_{\bar q}+{\vec L} $, where the ${\vec
s}_{\bar q}$ is the spin of the light antiquark $\bar q$ and the
${\vec L}$ is the orbital angular momentum of the light degrees of
freedom \cite{RevWise,RevNeubert}. In the quark models, we usually
use the $n$ to denote the radial quantum number. In the case $n=1$,
for $L=0$, the doublet $(P,P^*)$ have the spin-parity
$J^P_{s_\ell}=(0^-,1^-)_{\frac{1}{2}}$; $L=1$,
 the two doublets $(P^*_0,P^{\prime}_1)$ and $(P_1,P^*_2)$ have the spin-parity
$J^P_{s_\ell}=(0^+,1^+)_{\frac{1}{2}}$ and $(1^+,2^+)_{\frac{3}{2}}$
respectively; $L=2$, the  two doublets $(P^*_1,P_2)$ and
$(P_2^{\prime *},P_3)$ have the spin-parity
$J^P_{s_\ell}=(1^-,2^-)_{\frac{3}{2}}$ and $(2^-,3^-)_{\frac{5}{2}}$
respectively; where the superscript $P$ denotes the parity. The
$n=2,3,4, \cdots$ states are clarified by the  analogous  doublets,
for example, $n=2$, $L=0$, the  doublet $(P^{\prime},P^{*\prime})$
have the spin-parity $J^P_{s_\ell}=(0^-,1^-)_{\frac{1}{2}}$; $n=2$,
$L=1$, the two doublets $(P^{*\prime}_0,{P^{\prime}_1}^\prime)$ and
$({P_1}^{\prime},P^{*\prime}_2)$ have the spin-parity
$J^P_{s_\ell}=(0^+,1^+)_{\frac{1}{2}}$ and $(1^+,2^+)_{\frac{3}{2}}$
respectively.

 The helicity distributions favor identifying the $D^0(2550)$ as the
$0^-$ state,  the $D^0(2600)$ as the $1^-$, $2^+$, $3^-$ state, and
the  $D^0(2750)$ as the $1^+$, $2^-$ state \cite{Babar2010}. From
the Review of Particle Physics
 \cite{PDG}, we can see that  only six low-lying
states, $D$, $D^*$, $D_0(2400)$, $D_1(2430)$, $D_1(2420)$ and
$D_2(2460)$ are established, while the  $2S$ and $1D$ states are
still absent. The newly observed charmed mesons $D(2550)$,
$D(2600)$, $D(2750)$ and $D(2760)$ may be tentatively  identified as
the missing $2S$ and $1D$ states.

The mass is a fundamental parameter in describing a hadron, in Table
2,  we present the predictions  from some theoretical models,  such
as the relativized quark model based on a universal
one-gluon-exchange-plus-linear-confinement potential
 \cite{GI}, the semirelativistic quark potential model \cite{MMS}, the relativistic quark model includes the leading
order $1/M_h$ corrections \cite{PE}, the QCD-motivated relativistic
quark model based on the quasipotential approach \cite{EFG},  for
comparison. From the Table, we can see that the masses of the
$D(2550)$,  $D(2600)$ and  $D(2750)$, $D(2760)$ lie in the regions
of $2S$ and $1D$ states,  respectively.

\begin{table}
\begin{center}
\begin{tabular}{|c|c|c|c|c|c|c }\hline\hline
 & Mass [MeV] & Width [MeV] &  Decay channel   \\ \hline
 $D^0(2550)$ & $2539.4 \pm 4.5 \pm 6.8$ & $130\pm12 \pm13$ &$D^{*+}\pi^-$   \\ \hline
 $D^0(2600)$ & $2608.7\pm 2.4\pm 2.5$ & $93\pm 6\pm13$ &$D^+\pi^-$,\,$D^{*+}\pi^-$ \\ \hline
 $D^0(2750)$ & $2752.4\pm 1.7\pm 2.7$ & $71\pm6\pm11$ &$D^{*+}\pi^-$   \\ \hline
 $D^0(2760)$ & $2763.3\pm 2.3\pm 2.3$ & $60.9\pm5.1\pm3.6$ &$D^{+}\pi^-$       \\ \hline
 $D^+(2600)$ & $2621.3\pm 3.7\pm 4.2$ & $93$ &$D^0\pi^+$ \\ \hline
 $D^+(2760)$ & $2769.7\pm 3.8\pm 1.5$ & $60.9 $ &$D^0\pi^+$    \\ \hline\hline
\end{tabular}
\end{center}
\caption{The experimental results from the Babar collaboration.}
\end{table}

\begin{table}
\begin{center}
\begin{tabular}{|c|c|c|c|c|c|c|c|}\hline\hline
    & $n\,L\,s_\ell\,J^P$& Experiment \cite{Babar2010,PDG} &GI \cite{GI} & MMS \cite{MMS} & PE \cite{PE} & EFG \cite{EFG}\\ \hline
  $D$  & $1\,S\,\frac{1}{2}\,0^-$ & 1867&  1880& 1869& 1868& 1871\\ \hline
  $D^*$ & $1\,S\,\frac{1}{2}\,1^-$ & 2008 &2040 &2011& 2005& 2010\\ \hline
   $D^*_0$ & $1\,P\,\frac{1}{2}\,0^+$ & 2400 &2400 &2283& 2377& 2406\\ \hline
   $D^{\prime}_1$ & $1\,P\,\frac{1}{2}\,1^+$ & 2427& 2490& 2421& 2490& 2469\\ \hline
   $D_1$ & $1\,P\,\frac{3}{2}\,1^+$ & 2420& 2440& 2425& 2417& 2426\\ \hline
   $D_2^*$ & $1\,P\,\frac{3}{2}\,2^+$ & 2460 &2500& 2468& 2460& 2460\\ \hline
   $D_1^*$ & $1\,D\,\frac{3}{2}\,1^-$ & ?\,2763 &2820 & 2762 & 2795 &2788  \\ \hline
   $D_2$ & $1\,D\,\frac{3}{2}\,2^-$ & ?\,2752 & &  2800&  2833&2850  \\ \hline
   $D_2^{\prime*}$ & $1\,D\,\frac{5}{2}\,2^-$ & ?\,2752  & &  & 2775 & 2806 \\ \hline
   $D_3$ & $1\,D\,\frac{5}{2}\,3^-$ & ?\,2763 & 2830&  & 2799 &2863  \\ \hline
   $D^{\prime}$  & $2\,S\,\frac{1}{2}\,0^-$ & ?\,2539 & 2580& & 2589& 2581\\ \hline
   ${D^*}^{\prime}$  & $2\,S\,\frac{1}{2}\,1^-$ & ?\,2609& 2640& & 2692& 2632\\ \hline
  \hline
\end{tabular}
\end{center}
\caption{ The masses of the charmed mesons from different quark
models compared with experimental data, and the possible
identifications of the newly observed charmed mesons.  }
\end{table}

In Ref.\cite{Sun1008},  Sun et al study the strong decays of the
$D(2550)$, $D(2600)$ and $D(2760)$  in the $^3P_0$ model, and
identify the $D(2600)$ as a mixture of the $2^3S_1-1^3D_1$ states
and the $D(2760)$ as either the orthogonal partner of the $D(2600)$
or the $1^3D_3$ state.  In Ref.\cite{Zhong1009}, Zhong studies the
strong decays of the  $D(2550)$, $D(2600)$ and $D(2760)$ in a chiral
quark model, and identifies the $D(2760)$  as the $1^3D_3$ state and
the $D(2600)$ as the low-mass mixing  state of the $1^3D_1-2^3S_1$
states.

In this work, we study the strong decays  of the newly observed
charmed mesons with  the heavy quark effective theory in the leading
order approximation to distinguish the different identifications.
There have been several works using the heavy quark effective theory
to identify the excited $D_s$ mesons, such as the $D_s(3040)$,
$D_s(2700)$, $D_s(2860)$
\cite{Colangelo1001,Colangelo0710,Colangelo0607,Colangelo0511}.

The article is arranged as follows:  we study  the   strong decays
of the newly observed charmed mesons   with the heavy quark
effective theory in Sect.2; in Sect.3, we present the
 numerical results and discussions; and Sect.4 is reserved for our
conclusions.

\section{ The strong  decays with the heavy quark effective theory }

In the heavy quark effective theory,   the  spin doublets can be
described by the effective super-fields $H_a$, $S_a$,  $T_a$, $X_a$
and $Y_a$,  respectively \cite{Falk1992},
\begin{eqnarray}
H_a & =& \frac{1+{\rlap{v}/}}{2}\left\{P_{a\mu}^*\gamma^\mu-P_a\gamma_5\right\} \, ,   \nonumber  \\
S_a &=& \frac{1+{\rlap{v}/}}{2} \left\{P_{1a}^{\prime \mu}\gamma_\mu\gamma_5-P_{0a}^*\right\}  \, , \nonumber \\
T_a^\mu &=&\frac{1+{\rlap{v}/}}{2} \left\{ P^{\mu\nu}_{2a}\gamma_\nu
-P_{1a\nu} \sqrt{3 \over 2} \gamma_5 \left[ g^{\mu \nu}-{\gamma^\nu
(\gamma^\mu-v^\mu) \over 3} \right]\right\}\, ,   \nonumber\\
X_a^\mu &=&\frac{1+{\rlap{v}/}}{2} \Bigg\{ P^{*\mu\nu}_{2a}
\gamma_5\gamma_\nu -P^{\prime*}_{1a\nu} \sqrt{3 \over 2} \left[ g^{\mu \nu}-{\gamma^\nu (\gamma^\mu-v^\mu) \over 3}  \right]\Bigg\} \, , \nonumber  \\
Y_a^{ \mu\nu} &=&\frac{1+{\rlap{v}/}}{2} \left\{
P^{\mu\nu\sigma}_{3a} \gamma_\sigma -P^{\prime*\alpha\beta}_{2a}
\sqrt{5 \over 3} \gamma_5 \left[ g^\mu_\alpha g^\nu_\beta -
{\gamma_\alpha g^\nu_\beta (\gamma^\mu-v^\mu) \over 5}
 - {\gamma_\beta g^\mu_\alpha (\gamma^\nu-v^\nu) \over 5}  \right]
\right\}\, ,
\end{eqnarray}
where the  heavy field operators  contain a factor $\sqrt{M_P}$ and
have dimension of mass $\frac{3}{2}$. The ground state and radial
excited state heavy mesons with the same heavy flavor have the same
spin, parity, time-reversal and charge conjunction properties except
for the masses, and can be denoted by  the super-fields: $H_a$,
$H_a'$, $H_a''$, $\cdots$; $S_a$, $S_a'$, $S_a''$, $\cdots$; $T_a$,
$T_a'$, $T_a''$, $\cdots$; etc, where the superscripts  $\prime$,
$\prime\prime$ and $\prime\prime\prime$ denote the first, the second
and the third radial excited states, respectively. With a simple
replacement of the components $P_a$, $P_a^*$, $P_{0a}^*$, $\cdots$
to the corresponding radial excited states ${P_a}^\prime$,
${P_a^{*}}^\prime$, ${P_{0a}^{*}}^\prime$, $\cdots$, we can obtain
the corresponding super-fields $H_a^\prime$, $S_a^\prime$, $\cdots$.

 The light pseudoscalar mesons are described by the fields
 $\displaystyle \xi=e^{i {\cal M} \over
f_\pi}$, where
\begin{equation}
{\cal M}= \left(\begin{array}{ccc}
\sqrt{\frac{1}{2}}\pi^0+\sqrt{\frac{1}{6}}\eta & \pi^+ & K^+\nonumber\\
\pi^- & -\sqrt{\frac{1}{2}}\pi^0+\sqrt{\frac{1}{6}}\eta & K^0\\
K^- & {\bar K}^0 &-\sqrt{\frac{2}{3}}\eta
\end{array}\right) \, .
\end{equation}

At the leading order, the heavy meson chiral Lagrangians  ${\cal
L}_H$, ${\cal L}_S$, ${\cal L}_T$, ${\cal L}_X$, ${\cal L}_Y$  for
the strong decays to $D^{(*)}\pi$, $D^{(*)}\eta$ and $D_s^{(*)}K$
are written as \cite{HL-1,HL-2,HL-3,HL-4,PRT1997}:
\begin{eqnarray}
{\cal L}_H &=& \,  g_H {\rm Tr} \left\{{\bar H}_a H_b \gamma_\mu\gamma_5 {\cal A}_{ba}^\mu \right\} \, ,\nonumber \\
{\cal L}_S &=& \,  g_S {\rm Tr} \left\{{\bar H}_a S_b \gamma_\mu \gamma_5 {\cal A}_{ba}^\mu \right\}\, + \, h.c. \, , \nonumber \\
{\cal L}_T &=&  {g_T \over \Lambda_\chi}{\rm Tr}\left\{{\bar H}_a
T^\mu_b (i D_\mu {\not\! {\cal A}  }+i{\not\! D  }
 { \cal A}_\mu)_{ba} \gamma_5\right\} + h.c.  \, , \nonumber \\
{\cal L}_X &=&  {g_X \over \Lambda_\chi}{\rm Tr}\left\{{\bar H}_a
X^\mu_b(i D_\mu {\not\! {\cal A}  }+i{\not\! D  } { \cal A}_\mu)_{ba} \gamma_5\right\} + h.c.  \, ,\nonumber   \\
{\cal L}_{Y} &=&  {1 \over {\Lambda_{\chi}^2}}{\rm Tr}\left\{ {\bar
H}_a Y^{\mu \nu}_b \left[k_1 \{D_\mu, D_\nu\} {\cal A}_\lambda
 + k_2 (D_\mu D_\lambda { \cal A}_\nu + D_\nu D_\lambda { \cal
A}_\mu)\right]_{ba}  \gamma^\lambda \gamma_5\right\} + h.c. \, ,
  \end{eqnarray}
where
\begin{eqnarray}
{\cal D}_{\mu}&=&\partial_\mu+{\cal V}_{\mu} \, , \nonumber \\
 {\cal V}_{\mu }&=&\frac{1}{2}\left(\xi^\dagger\partial_\mu \xi+\xi\partial_\mu \xi^\dagger\right)\, , \nonumber \\
 {\cal A}_{\mu }&=&\frac{1}{2}\left(\xi^\dagger\partial_\mu \xi-\xi\partial_\mu  \xi^\dagger\right)\,  ,
\end{eqnarray}
$\Lambda_\chi$ is the chiral symmetry-breaking scale and taken as
$\Lambda_\chi = 1 \, $ GeV \cite{Colangelo1001}, the strong coupling
constants  $g_H$, $g_S$, $g_T$, $g_X$ and $g_Y=(k_1+k_2)$ can be
fitted  phenomenologically if there are enough experimental data.
The subscript indexes $H$, $S$, $T$, $X$ and $Y$ denote the
interactions between the super-field $H$ and the super-fields $H$,
$S$, $T$, $X$ and $Y$, respectively. We have smeared the
superscripts $\prime$, $\prime\prime$, $\prime\prime\prime$,
$\cdots$ for simplicity, the notation $g_H$ denotes the strong
coupling constants in the vertexes $HH{\cal A}$, $H'H{\cal A}$,
$H'H'{\cal A}$, $H''H{\cal A}$, $\cdots$, the notations $g_S$,
$g_T$, $g_X$ and $g_Y$ should be  understood in the same way. In
this article, we intend to study the ratios among different decay
channels, the strong coupling constants are canceled out with each
other, and cannot lead to confusion.

From the heavy meson chiral Lagrangians  ${\cal L}_H$, ${\cal L}_S$,
${\cal L}_T$, ${\cal L}_X$, ${\cal L}_Y$, we can obtain the  widths
$\Gamma$ for the strong decays to $D^{(*)}\pi$, $D^{(*)}\eta$ and
$D_s^{(*)}K$ easily,
\begin{eqnarray}
\Gamma&=&\frac{p_{cm}}{8\pi M^2 } |T|^2\, ,
\end{eqnarray}
where the $T$ denotes the scattering amplitudes, the $p_{cm}$ is the
momentum of the final states in the center of mass coordinate.

In calculations, we take the approximation ${\cal{A}}_\mu\approx
i\frac{\partial_\mu {\cal{M}}}{f_{\pi}} $.  In the case that  the
light pseudoscalar meson momenta are not very small, we should  add
other terms and introduce new unknown coupling constants.
Furthermore, the flavor and spin violation corrections of order
$\mathcal {O}(1/m_Q)$ to the heavy quark limit may be sizable, again
we should  introduce  new unknown coupling constants, which will not
necessarily  canceled  out in the ratios of the decay widths.  We
cannot estimate  the role and the size of such corrections on
general grounds, however,  we expect that they would not be larger
than (or as large as) the leading order contributions.

\section{Numerical Results}
The input parameters are taken  from the particle data group
$M_{\pi^+}=139.57\,\rm{MeV}$, $M_{\pi^0}=134.9766\,\rm{MeV}$,
$M_{K^+}=493.677\,\rm{MeV}$, $M_{\eta}=547.853\,\rm{MeV}$,
$M_{D^+}=1869.60\,\rm{MeV}$, $M_{D^0}=1864.83\,\rm{MeV}$,
$M_{D_s^+}=1968.47\,\rm{MeV}$, $M_{D^{*+}}=2010.25\,\rm{MeV}$,
$M_{D^{*0}}=2006.96\,\rm{MeV}$, $M_{D_s^{*+}}=2112.3\,\rm{MeV}$,
 $M_{D(2460)}=2460.1\,\rm{MeV}$ \cite{PDG}.

 The numerical values for the widths  of the strong
 decays
 \begin{eqnarray}
 D_2^* &\to& D^{*+}\pi^-, \,D^{+}\pi^- \, , \nonumber \\
  D^{\prime}  & \to &D^{*+}\pi^- , \, D^{*0}\pi^0 \, , \nonumber \\
 {D^*}^{\prime} (D_1^*,\,D_2,\,D_3)   & \to & D^{*+}\pi^-, \, D^{+}\pi^-, \, D^{*0}\pi^0, \, D^{0}\pi^0,
 \, D^{*0}\eta, \, D^{0}\eta, \,D^{*+}_sK^-, \,D_s^{+} K^-\, ,
 \end{eqnarray}
  are presented in Tables 3-4.

In Table 5, we present the experimental data for the ratio
$\frac{\Gamma(D^{+}\pi^-)}{\Gamma(D^{*+}\pi^-)}$ of the well
established meson $D_2^*(2460)$ from the Babar \cite{Babar2010},
CLEO \cite{CLEO1994,CLEO1990}, ARGUS \cite{ARGUS1989}, and ZEUS
\cite{ZEUS2009} collaborations,  the prediction $2.30$ from the
heavy quark effective theory in the leading order approximation is
in excellent agreement with the average experimental value $2.35$.
Compared with the experimental data from the Babar collaboration
$\frac{\Gamma(D^{+}\pi^-)}{\Gamma(D^{*+}\pi^-)}=1.47\pm0.03\pm0.16$
\cite{Babar2010}, the heavy quark effective theory in the leading
order approximation leads to a larger ratio.

The total decay widths of the $(D(2550),D(2600))$ with the
spin-parity $(0^-,1^-)_{\frac{1}{2}}$ are
$\Gamma_{D^{\prime}}\approx 1.7g_H^2\,\rm{GeV}$ and
$\Gamma_{D^{*\prime}}\approx 2.0g_H^2\,\rm{GeV}$, the ratio
$\frac{\Gamma_{D^{\prime}}}{\Gamma_{D^{*\prime}}}\approx0.85$, which
is smaller than the experimental data
$\frac{\Gamma_{D^{\prime}}}{\Gamma_{D^{*\prime}}}=1.40$, where we
have used the central values of the widths
$\Gamma_{D^{\prime}}\approx (130\pm12\pm13)\,\rm{MeV}$ and
$\Gamma_{D^{*\prime}}= (93\pm6\pm13)\,\rm{MeV}$ from the Babar
collaboration \cite{Babar2010}. For the charmed mesons, the leading
power flavor and spin violation corrections (of order $\mathcal
{O}(1/m_Q)$) to the heavy quark limit may be sizable, we have to
introduce new unknown coupling constants,  the discrepancy may be
smeared with the optimal parameters, furthermore,  more precise
measurements are needed to make a reliable comparison. In the case
of the  ratio $\frac{\Gamma_{D_1}}{\Gamma_{D_2^*}}$,  the prediction
$0.30$ from the heavy quark effective theory in the leading order
approximation is also smaller than  the experimental data $0.48$
from the Review of Particle Physics \cite{PDG}, if  the leading
power spin corrections
 to the heavy quark limit are taken into account,
the discrepancy can be smeared \cite{Falk1996}.

The ratio $\frac{\Gamma(D^{*\prime}\to
D^{+}\pi^-)}{\Gamma(D^{*\prime}\to D^{*+}\pi^-)}=0.82$ from the
heavy quark effective theory in the leading order approximation  is
larger than  the experimental data $0.32\pm0.02\pm0.09 $ from the
Babar  collaboration \cite{Babar2010}, just like in the case of the
ratio $\frac{\Gamma(D^*_2 \to D^{+}\pi^-)}{\Gamma(D^*_2 \to
D^{*+}\pi^-)}$, and again  more precise measurements are needed to
make a reliable comparison. The strong coupling constants
$g_{D^*D\pi}$ and $g_{D^*D^*\pi}$ receive sizable  contributions
from the flavor and spin violation corrections
\cite{PRT1997,Grinstein1995}, in the present case, the strong
coupling constants $g_{D^{*\prime} D\pi}$ and $g_{D^{* \prime}
D^*\pi}$ also receive the flavor and spin violation corrections
besides the leading order strong coupling constant $g_H$, which
maybe account for the discrepancy.  We can tentatively identify the
$(D(2550),D(2600))$ as the doublet $(0^-,1^-)_{\frac{1}{2}}$ with
$n=2$.

The existing theoretical estimations for the strong coupling
constant $g_H$ among the ground state heavy mesons ($n=1$) vary  in
a large range $g_H=0.1-0.6$, it is difficult to select the ideal
value (one can consult Ref.\cite{Wang2007} for more literatures), we
usually use the value determined  from the precise experimental data
on the decay $D^{*+} \to D^0 \pi^+$ from the CLEO collaboration
\cite{CLEO-gH1,CLEO-gH2}. In the present case, the strong coupling
constants involve the radial excited $S$-wave heavy mesons and
ground state $D$-wave heavy mesons, therefor the situation is more
involved, and it is impossible to determine the revelent parameters
with the heavy quark effective theory itself without enough
experimental data. The theoretical works focus on the strong
coupling constants $g_H$, $g_S$, $g_T$ of the ground state $S$-wave
and $P$-wave heavy mesons (one can consult
Refs.\cite{PRT1997,Wang2007,Wang2006} for more literatures), while
the works  on the strong coupling constants $g_H$, $g_S$, $g_T$ of
the radial excited $S$-wave and $P$-wave heavy mesons and   $g_X$,
$g_Y$ of the ground state $D$-wave heavy mesons are rare due to lack
experimental data \cite{Zhu1003}. In this article, we take the
strong coupling constants $g_H$, $g_T$, $g_X$ and $g_Y$ as unknown
parameters, and prefer the ratios of the decay widths in different
channels to compare with the experimental data.

From Table 4, we can see that if we identify the $(D(2760),D(2750))$
as the doublet $(1^-,2^-)_{\frac{3}{2}}$ with $n=1$, the ratio
$\frac{\Gamma(D_1^*\to D^{+}\pi^-)}{\Gamma(D_2\to
D^{*+}\pi^-)}=4.07$ from the leading order heavy quark effective
theory   deviates  from the experimental data $0.42\pm0.05\pm0.11$
greatly \cite{Babar2010}\footnote{We take the approximation
$\Gamma_{D(2760)}=\Gamma_{D(2750)}$.}, which requires the flavor and
spin violation corrections depressed by the inverse heavy quark mass
$1/m_Q$ are as large as the leading order contributions and have
opposite sign, it is impossible, as the heavy quark effective theory
has given  many successful descriptions of the hadron properties
\cite{RevWise,RevNeubert,PRT1997}. On the other hand, if we identify
the $(D(2750),D(2760))$ as the doublet $(2^-,3^-)_{\frac{5}{2}}$
with $n=1$, the deviation of the ratio $\frac{\Gamma(D_3\to
D^{+}\pi^-)}{\Gamma(D_2^{\prime*}\to D^{*+}\pi^-)}=0.80$ from the
upper bound of the experimental data $0.42\pm0.05\pm0.11$ is not
large \cite{Babar2010}, the contributions from the flavor and spin
violation corrections maybe smear the discrepancy.

We also explore the possible identification of  the $D(2760)$ and
$D(2750)$ as the same $3^-$ state with $n=1$, i.e. they are the
$D_3$ state,  the ratio $\frac{ \Gamma(D_3\to
D^{+}\pi^-)}{\Gamma(D_3\to D^{*+}\pi^-)}=1.94$ from the heavy quark
effective theory in the leading order approximation is too large
compared with the experimental data $\frac{\Gamma\left(D(2760)^0\to
D^+\pi^-\right)}{\Gamma\left(D(2750)^0\to
D^{*+}\pi^-\right)}=0.42\pm0.05\pm0.11$ \cite{Babar2010}, which
again requires the flavor and spin violation corrections depressed
by the inverse heavy quark mass $1/m_Q$ are as large as the leading
order contributions and have opposite sign, such an identification
is disfavored. On the other hand, the helicity distribution
 disfavors identifying the $D(2750)$ as the $3^-$ state
\cite{Babar2010}.  We can tentatively identify the
$(D(2750),D(2760))$ as the doublet $(2^-,3^-)_{\frac{5}{2}}$ with
$n=1$.

In this article, we also present the widths for the $D_s^{(*)}K$ and
$D^{(*)}\eta$ decays, where the strong coupling constants are
retained, the predictions can be confronted with the experiential
data in the future at the BESIII, KEK-B, RHIC, $\rm{\bar{P}ANDA}$
and LHCb.

\begin{table}
\begin{center}
\begin{tabular}{|c|c|c|c|c|c| }\hline\hline
    & $n\,L\,s_\ell\,J^P$& Mass [MeV] &Decay channels & Width [GeV]  \\ \hline
   $D_2^*$ & $1\,P\,\frac{3}{2}\,2^+$ & 2460.1 & $D^{*+}\pi^-$; $D^{+}\pi^-$& $0.0543879g_T^2$; $0.124928g_T^2$ \\ \hline
   $D^{\prime}$  & $2\,S\,\frac{1}{2}\,0^-$ & ?\,2539.4 &$D^{*+}\pi^-$; $D^{*0}\pi^0$& $1.13557g_H^2$; $0.583137g_H^2$  \\ \hline

   ${D^*}^{\prime}$  & $2\,S\,\frac{1}{2}\,1^-$ & ?\,2608.7& $D^{*+}\pi^-$; $D^{+}\pi^-$& $0.66068g_H^2$; $0.54317g_H^2$\\
         &  &  & $D^{*+}_sK^-$; $D_s^{+} K^-$& $0.000518592g_H^2$; $0.106459g_H^2$\\
        &  & & $D^{*0}\pi^0$; $D^{0}\pi^0$& $0.336747g_H^2$; $0.276487g_H^2$\\
        &  & & $D^{*0}\eta$; $D^{0}\eta$& $0.00841286 g_H^2$; $0.029364g_H^2$\\  \hline

   $D_1^*$ & $1\,D\,\frac{3}{2}\,1^-$ & ?\,2763.3 & $D^{*+}\pi^-$; $D^{+}\pi^-$ & $0.339606g_X^2$; $5.19392 g_X^2$  \\
         &  &  & $D_s^{*+}K^-$; $D_s^{+}K^-$  & $0.0632191g_X^2$; $1.86912 g_X^2$  \\
        &   &   & $D^{*0}\pi^0$; $D^{0}\pi^0$ & $0.173223g_X^2$; $2.65247 g_X^2$  \\
        &   &   & $D^{*0}\eta$; $D^{0}\eta$ & $0.0226441g_X^2$; $0.508904 g_X^2$  \\\hline

   $D_2$ & $1\,D\,\frac{3}{2}\,2^-$ & ?\,2752.4 &  $D^{*+}\pi^-$; $D^{+}\pi^-$ & $1.27691 g_X^2$; 0  \\
        &  &  &  $D_s^{*+}K^-$; $D_s^{+}K^-$  & $0.180643 g_X^2$; 0  \\
          &  &  &  $D^{*0}\pi^0$; $D^{0}\pi^0$ & $0.653307 g_X^2$; 0  \\
          &  &  &  $D^{*0}\eta$; $D^{0}\eta$ & $0.069308 g_X^2$; 0  \\\hline

   $D_2^{\prime*}$ & $1\,D\,\frac{5}{2}\,2^-$ & ?\,2752.4  & $D^{*+}\pi^-$; $D^{+}\pi^-$ & $0.221226 g_Y^2$; 0   \\
        &  &   & $D_s^{*+}K^-$; $D_s^{+}K^-$ & $0.00413833g_Y^2$; 0   \\
        &  &   & $D^{*0}\pi^0$; $D^{0}\pi^0$ & $0.114719 g_Y^2$; 0   \\
        &  &   & $D^{*0}\eta$; $D^{0}\eta$ & $0.0027123 g_Y^2$; 0   \\ \hline

   $D_3$ & $1\,D\,\frac{5}{2}\,3^-$ & ?\,2763.3 & $D^{*+}\pi^-$; $D^{+}\pi^-$  & $0.0907266g_Y^2$; $0.176388g_Y^2$   \\
        &  & & $D_s^{*+}K^-$; $D_s^{+}K^-$& $0.00218128 g_Y^2$; $0.018115 g_Y^2$  \\
      &   &   & $D^{*0}\pi^0$; $D^{0}\pi^0$  & $0.0468994g_Y^2$; $0.0912646 g_Y^2$   \\
      &   &   & $D^{*0}\eta$; $D^{0}\eta$  & $0.00124089g_Y^2$; $0.00618076 g_Y^2$   \\ \hline
    \hline
\end{tabular}
\end{center}
\caption{ The strong decay widths of the newly observed charmed
mesons with possible identifications.   }
\end{table}

\begin{table}
\begin{center}
\begin{tabular}{|c|c|c|c| }\hline\hline
    & $n\,L\,s_\ell\,J^P$& Mass [MeV] & Ratio \\ \hline
   $D_2^*$ & $1\,P\,\frac{3}{2}\,2^+$ & 2460.1 & $\frac{\Gamma(D^{+}\pi^-)}{\Gamma(D^{*+}\pi^-)}=2.30$ \\ \hline

  ${D^*}^{\prime}$  & $2\,S\,\frac{1}{2}\,1^-$ & ?\,2608.7& $\frac{\Gamma(D^{+}\pi^-)}{\Gamma(D^{*+}\pi^-)}=0.82$;
  $\frac{\Gamma(D^{*0}\pi^0)}{\Gamma(D^{*+}\pi^-)}=0.51$; $\frac{\Gamma(D^{0}\pi^0)}{\Gamma(D^{*+}\pi^-)}=0.42$;\\
        &  & & $\frac{\Gamma(D_s^{+} K^-)}{\Gamma(D^{*+}\pi^-)}=0.16$; $\frac{\Gamma(D^{0}\eta)}{\Gamma(D^{*+}\pi^-)}=0.044$; $\frac{\Gamma(D^{*0}\eta)}{\Gamma(D^{*+}\pi^-)}=0.013$; \\
   &  & &  $\frac{\Gamma(D^{*+}_sK^-)}{\Gamma(D^{*+}\pi^-)}=0.001$ \\ \hline

   $D_1^*$ & $1\,D\,\frac{3}{2}\,1^-$ & ?\,2763.3 & $\frac{\Gamma(D^{+}\pi^-)}{\Gamma(D^{*+}\pi^-)}=15.29$; $\frac{\Gamma(D^{0}\pi^0)}{\Gamma(D^{*+}\pi^-)}=7.81$;  $\frac{\Gamma(D_s^{+}K^-)}{\Gamma(D^{*+}\pi^-)}=5.50$; \\
        &  &  & $\frac{\Gamma(D^{0}\eta)}{\Gamma(D^{*+}\pi^-)}=1.50$; $\frac{\Gamma(D^{*0}\pi^0)}{\Gamma(D^{*+}\pi^-)}=0.51$;  $\frac{\Gamma(D_s^{*+}K^-)}{\Gamma(D^{*+}\pi^-)}=0.19$; \\
         &  &  &  $\frac{\Gamma(D^{*0}\eta)}{\Gamma(D^{*+}\pi^-)}=0.067$ \\ \hline

   $D_2$ & $1\,D\,\frac{3}{2}\,2^-$ & ?\,2752.4 &  $\frac{\Gamma(D^{*0}\pi^0)}{\Gamma(D^{*+}\pi^-)}=0.51$; $\frac{\Gamma(D_s^{*+}K^-)}{\Gamma(D^{*+}\pi^-)}=0.14$;
           $\frac{\Gamma(D^{*0}\eta)}{\Gamma(D^{*+}\pi^-)}=0.054$  \\\hline

   $D_2^{\prime*}$ & $1\,D\,\frac{5}{2}\,2^-$ & ?\,2752.4  & $\frac{\Gamma(D^{*0}\pi^0)}{\Gamma(D^{*+}\pi^-)}=0.52$; $\frac{\Gamma(D_s^{*+}K^-)}{\Gamma(D^{*+}\pi^-)}=0.019$;
    $\frac{\Gamma(D^{*0}\eta)}{\Gamma(D^{*+}\pi^-)}=0.012$ \\    \hline

   $D_3$ & $1\,D\,\frac{5}{2}\,3^-$ & ?\,2763.3 & $\frac{\Gamma(D^{+}\pi^-)}{\Gamma(D^{*+}\pi^-)}=1.94$; $\frac{\Gamma(D^{0}\pi^0)}{\Gamma(D^{*+}\pi^-)}=1.01$;  $\frac{\Gamma(D^{0*}\pi^0)}{\Gamma(D^{*+}\pi^-)}=0.52$;  \\
     &   &   & $\frac{\Gamma(D_s^{+}K^-)}{\Gamma(D^{*+}\pi^-)}=0.20$;   $\frac{\Gamma(D^{0}\eta)}{\Gamma(D^{*+}\pi^-)}=0.068$; $\frac{\Gamma(D_s^{*+}K^-)}{\Gamma(D^{*+}\pi^-)}=0.024$; \\
        &  & &   $\frac{\Gamma(D^{0*}\eta)}{\Gamma(D^{*+}\pi^-)}=0.014$    \\ \hline

 $D_1^*$ & $1\,D\,\frac{3}{2}\,1^-$ & ?\,2763.3 &       \\
   $D_2$ & $1\,D\,\frac{3}{2}\,2^-$ & ?\,2752.4 &  $\frac{\Gamma(D_1^*\to D^{+}\pi^-)}{\Gamma(D_2\to D^{*+}\pi^-)}=4.07$    \\ \hline

$D_2^{\prime*}$ & $1\,D\,\frac{5}{2}\,2^-$ & ?\,2752.4  &    \\
   $D_3$ & $1\,D\,\frac{5}{2}\,3^-$ & ?\,2763.3 & $\frac{\Gamma(D_3^*\to D^{+}\pi^-)}{\Gamma(D_2^{\prime*}\to D^{*+}\pi^-)}=0.80$  \\ \hline
    \hline
\end{tabular}
\end{center}
\caption{ The ratios of the strong decay widths of the newly
observed charmed mesons with possible identifications.   }
\end{table}

\begin{table}
\begin{center}
\begin{tabular}{|c|c|c|c|c|c|c| }\hline\hline
   Babar&  CLEO& CLEO & ARGUS&ZEUS &This work\\ \hline
  $1.47\pm0.03\pm0.16$& $2.2\pm0.7\pm0.6$& $2.3\pm 0.8$ & $3.0\pm1.1\pm1.5$ &$2.8\pm 0.8^{+0.5}_{-0.6}$ &$2.30$\\ \hline
       \hline
\end{tabular}
\end{center}
\caption{ The ratio of $\frac{\Gamma\left(D^*_2(2460)^0\to
D^+\pi^-\right)}{\Gamma\left(D^*_2(2460)^0\to D^{*+}\pi^-\right)} $
from the experimental data compared with the prediction from the
leading order heavy quark effective theory. }
\end{table}

\section{Conclusion}
In this article, we study the strong decays of  the newly observed
charmed mesons $D(2550)$, $D(2600)$, $D(2750)$ and $D(2760)$ with
the heavy quark effective theory in the leading order approximation,
and tentatively identify the $(D(2550),D(2600))$ as the doublet
$(0^-,1^-)$ with $n=2$ and $(D(2750),D(2760))$ as the doublet
$(2^-,3^-)$ with  $n=1$, respectively. The identification of the
$D(2750)$ and $D(2760)$ as the same particle with $J^P=3^-$ is
disfavored. The other predictions can be confronted with the
experimental data in the future  at the BESIII, KEK-B, RHIC,
$\rm{\bar{P}ANDA}$ and LHCb.

\section*{Acknowledgment}
This  work is supported by National Natural Science Foundation of
China, Grant Numbers 10775051, 11075053, and Program for New Century
Excellent Talents in University, Grant Number NCET-07-0282, and the
Fundamental Research Funds for the Central Universities.

\end{document}